\documentclass[doublecol,linenumbers]{epl2}
\usepackage{graphics,graphicx,epsfig}
\usepackage{amssymb}
\usepackage{color}
\usepackage{comment}
\usepackage{subfigure}
\usepackage{hyperref}

\title{Temperature effect on $(2+1)$ experimental Kardar-Parisi-Zhang growth}
\shorttitle{Temperature effect on $(2+1)$ experimental KPZ growth} 

\author{R. A. L. Almeida\inst{1} \and S. O. Ferreira\inst{1} \and I. R. B. Ribeiro\inst{1,2} \and T. J. Oliveira\inst{1}}
\shortauthor{R. A. L. Almeida \etal}

\institute{                    
  \inst{1} Departamento de F\'isica, Universidade Federal de Vi\c cosa, 36570-000, Vi\c cosa, MG, Brazil\\
  \inst{2} Instituto Federal do Esp\'irito Santo, 29520-000, Alegre, ES, Brazil
}
\pacs{68.43.Hn}{Structure of assemblies of adsorbates (two- and three-dimensional clustering)}
\pacs{81.15.Aa}{Theory and models of film growth}
\pacs{05.40.-a}{Fluctuation phenomena, random processes, noise, and Brownian motion}

\abstract{We report on the effect of substrate temperature ($T$) on both local structure and long-wavelength fluctuations of polycrystalline CdTe thin films deposited on Si(001). {A strong $T$-dependent mound evolution is observed and explained in terms of the energy barrier to inter-grain diffusion at grain boundaries, as corroborated by Monte Carlo simulations}. This leads to transitions from uncorrelated growth to a crossover from random-to-correlated growth and transient anomalous scaling as $T$ increases. Due to these finite-time effects, {we were not able to determine the universality class of the system through the critical exponents}. Nevertheless, we demonstrate that this can be circumvented by analyzing height, roughness and maximal height distributions, which allow us to prove that CdTe grows {asymptotically} according to the Kardar-Parisi-Zhang (KPZ) equation in a broad range of $T$. More important, {one finds} positive (negative) velocity excess in the growth at low (high) $T$, indicating that it is possible to control the KPZ non-linearity by adjusting the temperature.}

\begin{document}

\maketitle


Thin films are the basis of the optoelectronic industry. Commonly, patterned/mounded interfaces are observed due to growth instabilities~\cite{evansrev} or polycrystallinity, where a complex growth involving intra- and inter-grain dynamics arise. It is well-known that size, texture and spatial distribution of these structures affect several thin-film properties that are crucial for applications in solar cells~\cite{solarcell}, spintronic devices~\cite{spin}, contact technology~\cite{contact} and many others. 

{At a coarse-grained level, the evolution of thin films and other growing interfaces is also a subject of broad interest, since they exhibit scaling invariance and universality}~\cite{barabasi,Krugrev}. For instance, the kinetic roughening of flame fronts~\cite{Marco}, turbulent phases in liquid crystals~\cite{Takeuchi}, colloidal particles deposited at the edges of evaporating drops~\cite{Yunker}, silica~\cite{Cuerno}, CdTe~\cite{Renan} and oligomer films~\cite{HHExper} have been shown to belong to the celebrated Kardar-Parisi-Zhang (KPZ)~\cite{KPZ} universality class (UC). Thus, despite their distinct microscopic nature, the interface of all these systems evolve asymptotically according to the KPZ equation~\cite{KPZ}:
 \begin{equation}
 \frac{\partial h (\textbf{x},t)}{\partial t} = \nu \nabla^{2} h + \frac{\lambda}{2} (\nabla h)^{2} + \eta(\textbf{x},t),
 \label{eqKPZ}
 \end{equation}
where $h(\textbf{x},t)$ is the height at substrate position $\textbf{x}$ and time $t$, and $\nu$, $\lambda$ and $\eta$ account, respectively, for surface tension, interfacial velocity excess and white noise.
  
Since the design of thin films with specific properties requires control of their growth, a natural question raises up: \textit{How do the growth parameters affect the local and long-wavelength dynamics of the system?} In fact, the effects of parameters such as substrate temperature \cite{Ferreira}, molecular flux \cite{Hamouda} and electric potential \cite{Huo} have already been studied. However, the (roughness) dynamic scaling analysis (DSA) performed there did not lead to any conclusion about the UC of those systems, possibly due to corrections to scaling/transient effects.

\begin{figure*}[!t]
\includegraphics*[width=17.0cm]{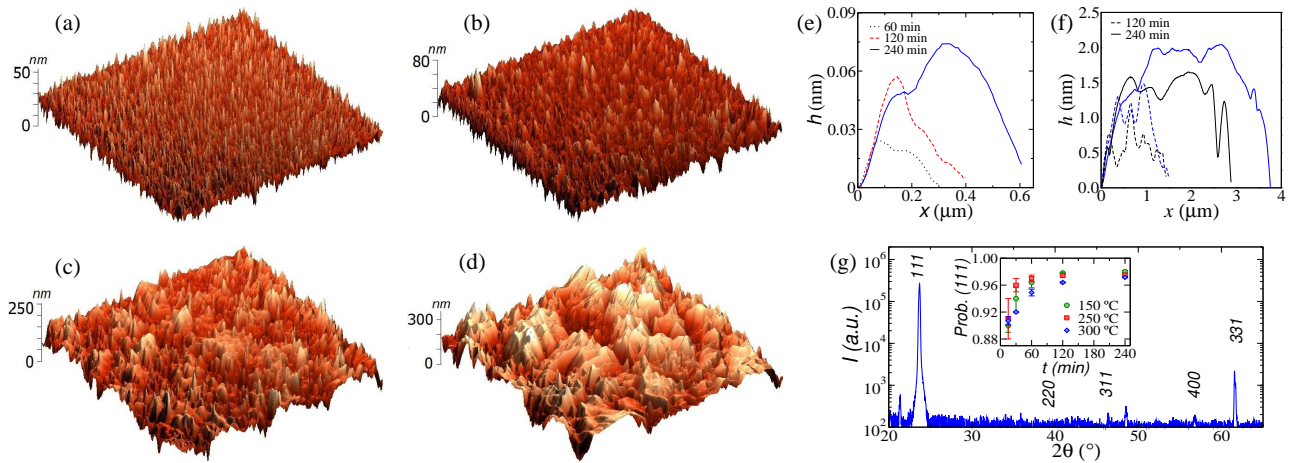}
\caption{$10 \mu$m $\times$ $10 \mu$m AFM images for films grown at 
$T=200\,^{\circ}\mathrm{C}$ [(a) and (b)] and $T=300\,^{\circ}\mathrm{C}$ [(c) and (d)] for $120$ (left) 
and 240 min (right). Cross sections of typical mounds for $T = 200$ and $300\,^{\circ}\mathrm{C}$ are shown in
(e) and (f), respectively. (g) XRD results for $T = 250\,^{\circ}\mathrm{C}$ and deposition time $t=240$min. 
The inset shows the coefficient of preferential growth in [111] direction as a function of growth time.}
\label{fig1}
\end{figure*}

In this Letter, we answer that question regarding the effect of  deposition temperature ($T$) on the growth of CdTe thin films, a very important material for the fabrication of detectors (of $\gamma$- and X-rays)~\cite{Ferreira}, solar cells~\cite{CLi}, ultra-fast optical sensors~\cite{Saba} and others~\cite{Manna}. The structure and morphology of films, grown by Hot Wall Epitaxy (HWE) at different $T$, have been characterized by X-ray diffraction (XRD) and atomic force microscopy (AFM). A complex mound evolution is observed and has been explained in terms of diffusion barriers at grain boundaries, as confirmed by Monte Carlo simulations. Although transient/crossover effects in the DSA have also been found here, through the study of several distributions, we prove that long-wavelength dynamics of the CdTe surfaces belong to the KPZ class in a broad range of $T$. The $T$-dependence of the parameters in Eq. \ref{eqKPZ} is also unveiled.


\section{Experimental methods}

CdTe (5N) was evaporated on Si(001) substrates by HWE, a well-controlled and highly reproducible growth technique \cite{Otero, Suela}. The experimental HWE setup is described in detail in Ref. \cite{Suela}. Substrate cleaning and growth conditions are the same reported in \cite{Renan}, however, temperatures $T=150$, $200$ and $300\,^{\circ}\mathrm{C}$ are also considered here. 
The HWE technique has been chosen because the growth apparatus is much simpler and the running costs are lower than those 
for Molecular Beam Epitaxy (MBE). The growth rate was determined ex-situ 
measuring the sample thickness with a ContourGT-K (Bruker) optical profiler. Surface topographies were measured in air using a Ntegra Prima (NT-MDT) SPM in contact mode with Si tips. Images of $10\mu m \times 10\mu m$ ($1024 \times 1024$ pixels) were carried out for 3-10 distinct regions near the film center. 
Crystallinity and texture features were investigated by X-ray diffraction (XRD) using a Bruker D8-Discover diffractometer.


\section{Local dynamic}

Figures \ref{fig1}a-d show typical AFM images for CdTe films grown at $T=200$ and $300\,^{\circ}\mathrm{C}$ for 120 and 240 min. The grained/mounded morphology expected for polycrystalline surfaces is observed, where for $T = 200\,^{\circ}\mathrm{C}$ conical grains with well-defined boundaries dominate the surface at short times (Fig. \ref{fig1}a). For the largest time available (Fig. \ref{fig1}b) one sees the presence of some coalesced/packed grains\footnote{{The large structures (mounds) can be formed by both coalescence of grains with the same crystallographic orientation and packing of grains with different orientations.}} carrying a multi-peaked form. There is a plenty of these structures at surfaces grown at $T = 300\,^{\circ}\mathrm{C}$ since short growth times (Fig \ref{fig1}c), and they keep growing to give place to large mounds separated by deep valleys (Fig \ref{fig1}d). These features are highlighted in Figs. \ref{fig1}e and \ref{fig1}f, where cross sections of characteristic superficial structures are shown. The polycrystalline environment is confirmed by the appearance of several peaks in the $\theta-2\theta$ XRD spectra. It is presented in Fig. \ref{fig1}g for $T=250\, ^{\circ}\mathrm{C}$, and similar spectra are found for all investigated $T$. Additionally, a strong (111) $T$-independent texture is revealed (see inset of  Fig. \ref{fig1}g), pointing out that (111) grains grow faster than the others and, upon {coalescence/packing}, cover the neighboring non-(111) ones. Previous studies suggest that this CdTe texture is also independent of the substrate \cite{Igor}.

\begin{figure}[!t]
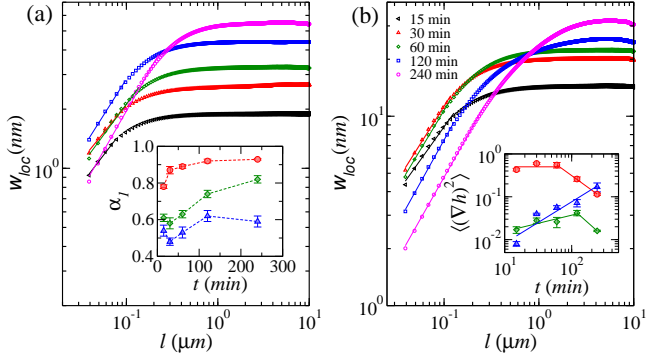

\includegraphics*[width=4.20cm]{Fig2a.eps}
\includegraphics*[width=4.20cm]{Fig2b.eps}
\caption{Local roughness $w_{loc}$ versus box size $l$ for (a) $T = 200$ and 
(b) $300\,^{\circ}\mathrm{C}$. Solid lines indicate linear fits used to extract the local exponent 
$\alpha_{1}$. In (a) and (b) insets show $\alpha_{1}$ and $\langle (\nabla h)^{2} \rangle$ against 
time, respectively, {for $T = 150\,^{\circ}\mathrm{C}$ (blue triangles), $200\,^{\circ}\mathrm{C}$ (green diamonds) and $300\,^{\circ}\mathrm{C}$ (red circles)}.}
\label{fig2}
\end{figure}

Figures \ref{fig2}a and \ref{fig2}b show the local roughness [$w_{loc}(l,t)$ - defined as the rms height fluctuation inside a box of lateral size $l$] versus $l$ for $T=200$ and $300\,^{\circ}\mathrm{C}$, respectively. Solid lines indicate linear fits used to extract the local exponent $\alpha_{1}$, defined by $w_{loc} \sim l^{\alpha_1}$~\cite{tiago3}. One notices that this exponent is measured for $l\lesssim 0.1 \mu$m, so that it characterizes the intra-mound morphology. As demonstrated in Ref. \cite{tiago3}, $\alpha_1$ decreases, as sharper are the mound shapes, from $\alpha_1 \approx 0.90$ (for rounded mounds) {down to} $\alpha_1 \approx 0.50$. Therefore, the exponents depicted in the inset of Fig. \ref{fig2}a indicate that, for a given $T$, the top of mounds becomes more rounded as time evolves. The same is seen when the time is fixed and $T$ increases, as corroborated in Figs. \ref{fig1}a-f. {Unfortunately, a crossover to the truly roughness exponent \cite{tiago3} is not observed in the local roughness (Figs. \ref{fig2}a-b).}

One may notice in Fig. \ref{fig2}a, at short-length scales ($l\lesssim 0.1 \mu$m), $w_{loc}$ increasing in time for $t\leqslant 120$ min. This is the hallmark of anomalous scaling~\cite{Anom}, but this ``anomaly'' is transient, since at large $t$ one sees $w_{loc}$ decreasing in time, leading to the standard Family-Vicsek scaling \cite{FVS}. A similar behavior was found in CdTe films grown at $T = 250\,^{\circ}\mathrm{C}$ \cite{Renan}. The time evolution of $w_{loc}$  at short-length scales is directly related to the spatially averaged squared local slopes $\left\langle (\nabla h)^2 \right\rangle$ at interface (see the inset of Fig. \ref{fig2}b).

The origin of these results can be understood as follows. Initially, CdTe grains evolve in the Volmer-Weber growth mode \cite{Sukarno}, and {as higher is $T$ larger are their widths} \cite{evansrev} (Fig. \ref{fig1}). Moreover, one found that the grain aspect ratio ($r \equiv$ height/width) also increases with $T$ at short times, possibly due to an unbalance between up- and downward diffusion at grain edges. As the initially isolated grains enlarge laterally, they collide forming grain boundaries (GBs), where defects are formed. {These defects give rise to an additional energy barrier $E_{GB}$ to diffuse toward these sites \cite{Tello_Gonzales}, as also suggested recently in the growth of CdTe/CdS films \cite{Kwon}. At low $T$, a small number of molecules overcomes this barrier and most of them aggregates inside the grain where they have arrived. This compels the grain height to increase faster than its width, leading $r$ and $\left\langle (\nabla h)^2 \right\rangle$ to increase (Figs. \ref{fig1}e and \ref{fig2}b). As time evolves, aggregations at GBs induces a relaxation, which diminishes the number of \textit{superficial} defects \cite{Sivananthan} and, consequently, the inter-grain diffusion becomes more active, since the $E_{GB}$ barrier disappear in those relaxed places. Thence, the coalescence/packing of grains becomes more operative and small grains give place to larger width structures, so that $r$ and $\left\langle (\nabla h)^2 \right\rangle$ start to decrease, exactly as observed for $T=200\,^{\circ}\mathrm{C}$ (Fig. \ref{fig2}b). For higher $T$, where the surface diffusion is more active, the relaxation process happens earlier as well as the decreasing in $r$ and $\left\langle (\nabla h)^2 \right\rangle$ (Figs. \ref{fig1}f, \ref{fig2}b and Ref. \cite{Renan}).}


\begin{figure}[!t]
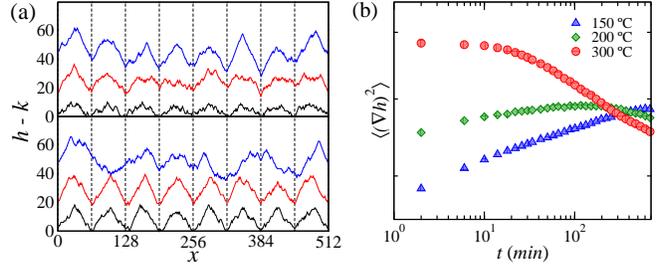

\includegraphics*[height=3.50cm]{Fig3a.eps}
\includegraphics*[height=3.50cm]{Fig3b.eps}
\caption{(a) Height profiles for $T=150$ (top) and $200\,^{\circ}\mathrm{C}$ 
(bottom) for $t=10$, 100 and 1000, and shifted by $k=10$, 80 and 960, respectively. The vertical 
lines indicate the location of the initial GBs. (b) Average squared local slope $\langle (\nabla h)^{2} \rangle$ versus time.}
\label{fig3}
\end{figure}

The reliability of the above reasoning is illustrated in a {very simplified} one-dimensional atomistic growth model. Since our interest is the coalescence process, the growth starts on a periodic array of pyramidal grains with the same width $\zeta$ and height $H$, for simplicity. A randomly deposited particle diffuses at surface until reaches a site $i$ satisfying the constraint $|h_i - h_{i\pm 1}| \leq 1$, where it permanently aggregates. Therefore, inside the grain, aggregation follows the conservative RSOS (restricted solid-on-solid) rule \cite{CRSOS}. However, at the GBs there is an energy barrier $E_{GB}$, so that a particle diffuses toward them with probability $P_{D}=e^{-E_{GB}/k_{B} T}$. Once a particle \textit{aggregates} at a given GB $i$, the barrier $E_{GB}$ at $i$ becomes null with probability $P_{R}=e^{-E_{R}/k_{B} T}$, in order to mimic the relaxation process. Figure \ref{fig3}a shows typical surface evolutions for $T = 150$ and $200\,^{\circ}\mathrm{
 C}$, with $E_{GB}=0.10$ eV, $E_{R}=0.30$ eV, $\zeta=64$ and $H=8$, 16 and 24 for $T=150$, 200 and $300\,^{\circ}\mathrm{C}$, respectively. For $T=150\,^{\circ}\mathrm{C}$ one observes grains with almost fixed widths and increasing heights. A similar behavior is found at short times for $T=200\,^{\circ}\mathrm{C}$ but, for long $t$, large mounds (formed by coalesced grains) appears. The same occurs for higher $T$. This qualitative agreement with the experiment is corroborated by the evolution of $\langle (\nabla h)^{2} \rangle$, displayed in Fig. \ref{fig3}b. Comparing these results with the experimental ones (inset of Fig. \ref{fig2}b), one can confirm that the interplay of GBs barrier relaxation and initial conditions (initial $r$ increasing with $T$), in fact, explains the CdTe/Si(001) local evolution. {Despite this agreement, we remark that this model does not captures all aspects of the microscopic dynamics of the system as well as of the complex packing of polycrystalline grains.}


\section{Coarse-grained dynamic} 

\begin{figure}[!t]
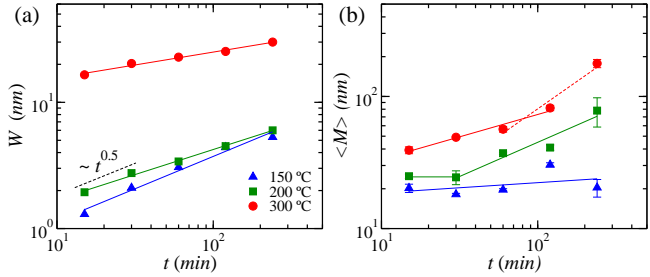

\includegraphics*[width=4.20cm]{Fig4a.eps}
\includegraphics*[width=4.20cm]{Fig4b.eps}
\caption{(a) Global roughness $W$ and (b) first zero of SSCF $\left\langle M\right\rangle $ versus time. The 
lines indicate linear fits used to extract the exponents $\beta$ [in (a)] and $1/z$ [in (b)].}
\label{fig4}
\end{figure}

Figure \ref{fig4}a presents the global roughness [$W(t) \equiv w_{loc}(l=L)$] versus time, which is expected to scale as $W \sim t^{\beta}$ \cite{barabasi,Krugrev}. From the linear fits in Fig. \ref{fig4}a, one obtains $\beta = 0.51(4)$, $0.41(5)$ and $0.21(5)$ for $T=150$, $200$ and $300\,^{\circ}\mathrm{C}$, respectively. This last value is consistent with the KPZ one ($\beta_{KPZ} \approx 0.24$), as was also found for $T=250\,^{\circ}\mathrm{C}$ \cite{Renan}. In turn, for $T=150\,^{\circ}\mathrm{C}$, the value is consistent with an uncorrelated growth (where $\beta=1/2$ \cite{barabasi,Krugrev}), whereas for $T=200\,^{\circ}\mathrm{C}$ the exponent does not correspond to any known UC, possibly due to crossover effects. Indeed, for short times, one finds an initial slope $\beta_{eff} \approx 1/2$ and a tendency of $\beta_{eff}$ to decrease in time.


From the first zero of the slope-slope correlation function (SSCF) $\Gamma(l,t) \equiv \left\langle \nabla h (\textbf{x},t) \nabla h (\textbf{x}+\textbf{l},t) \right\rangle$ one may estimate the average grain width $\left\langle M \right\rangle $ \cite{Siniscalco}. This quantity is expected to be of the same order of the correlation length $\xi(t)$, which scales as $\xi \sim t^{1/z}$, where $z$ is the dynamic exponent \cite{Krugrev}. For all $T$ analyzed here, $\Gamma(l,t)$ presented an oscillatory behavior similar to that found in Ref. \cite{Renan} for $T=250\,^{\circ}\mathrm{C}$. The values of $\left\langle M(t) \right\rangle$ extracted from those plots are depicted in Fig. \ref{fig4}b. For $T = 150\,^{\circ}\mathrm{C}$, one finds $\xi \sim \left\langle M \right\rangle \approx const$, confirming that the growth is uncorrelated. A similar behavior is found for $T = 200\,^{\circ}\mathrm{C}$ at short times. In contrast, for large $t$, a finite $z$ arises, namely $z=1.8(5)$, which agree with the KPZ value ($z_{KPZ} \approx 1.62$) within the error bar. Finally, for $T= 300\,^{\circ}\mathrm{C}$, linear fits at short and long times yields $z \approx 3.0$ and $z \approx 1.3$, respectively. However, in this case one can not ensure that $\xi \sim \left\langle M \right\rangle$, because small grains at the top of the multi-peaked mounds (see Fig. \ref{fig1}f) can make $\left\langle M \right\rangle$ smaller than $\xi$. This deposes against the reliability of a large $z$. Anyway, one notice that $\beta=0.21(5)$ and $z\approx 3.0$ are consistent with the Villain-Lai-Das Sarma (VLDS) class \cite{VLDS}, while a small $z$ suggests KPZ growth. Thus, at this point, based only on the DSA, one can not decide what is the UC of the films grown at $T = 200$ and $300\,^{\circ}\mathrm{C}$.


Beyond the scaling exponents, height distributions (HDs)\cite{Takeuchi,Yunker,HDsteor,Halpin,tiago2d}, 
squared local roughness distributions (SLRDs)~\cite{SLRDs, Paiva}, maximal relative height distributions (MRHDs) \cite{MRHDs} are also expected to be universal. {Indeed, the universality of these distributions, in $(2+1)$ KPZ class, has been experimentally demonstrated, by us, in the growth of CdTe/Si(001) at $T = 250\,^{\circ}\mathrm{C}$ \cite{Renan}}.
More recently, Halpin-Healy and Palasantzas have applied the same method to confirm KPZ growth in oligomer films~\cite{HHExper}. We remark that SLRDs and MRHDs are built by measuring the squared roughness ($w^2$) and the relative maximal height $m = h_{max} - \left\langle h \right\rangle $ into boxes of lateral size $l$ spanning the whole surface. The size $l$ must be larger than the pixel size and $l \ll \xi$\cite{Paiva}.

\begin{figure}[!t]
\includegraphics*[width=4.20cm]{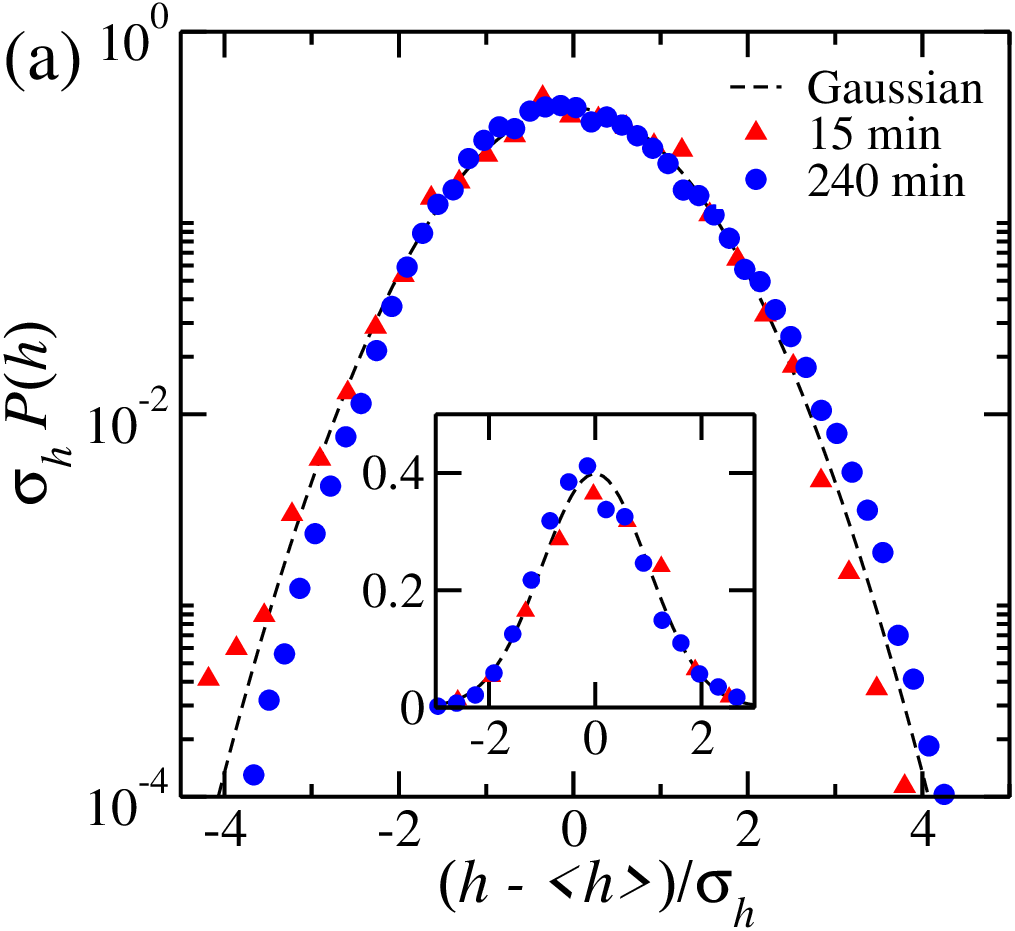}
\includegraphics*[width=4.20cm]{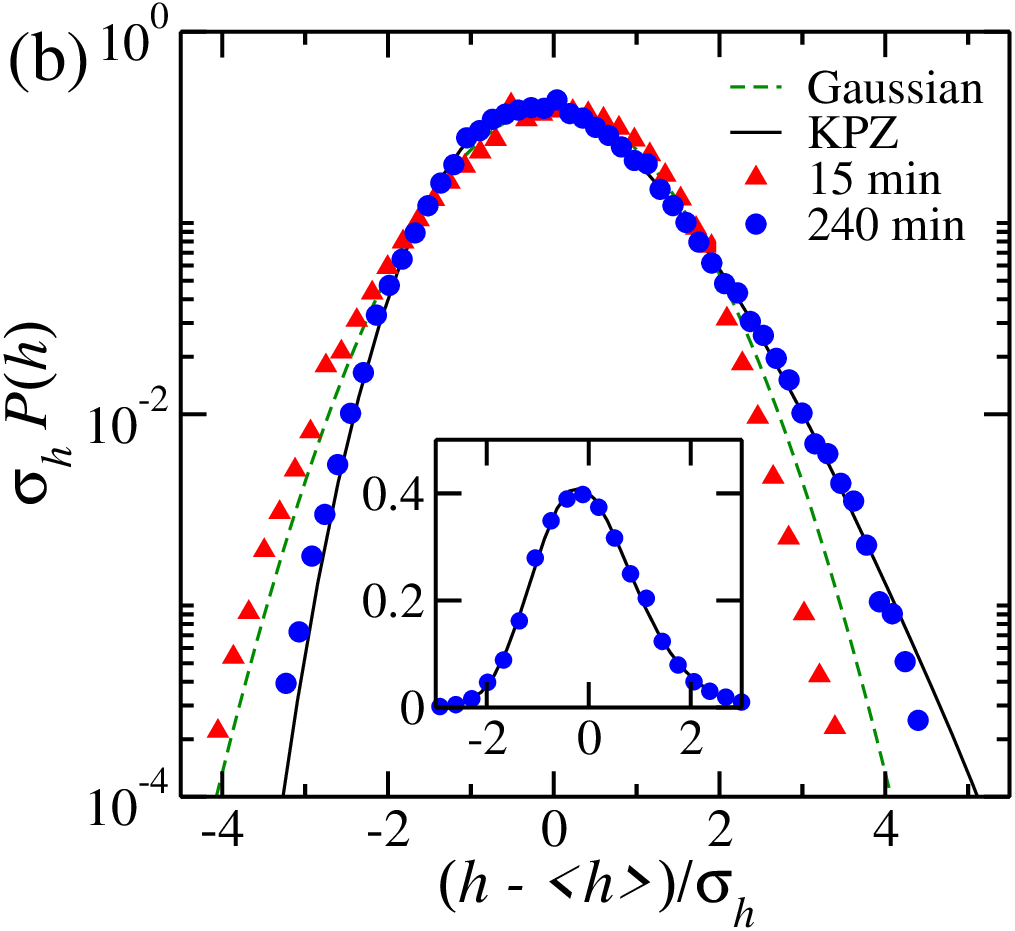}
\includegraphics*[width=4.20cm]{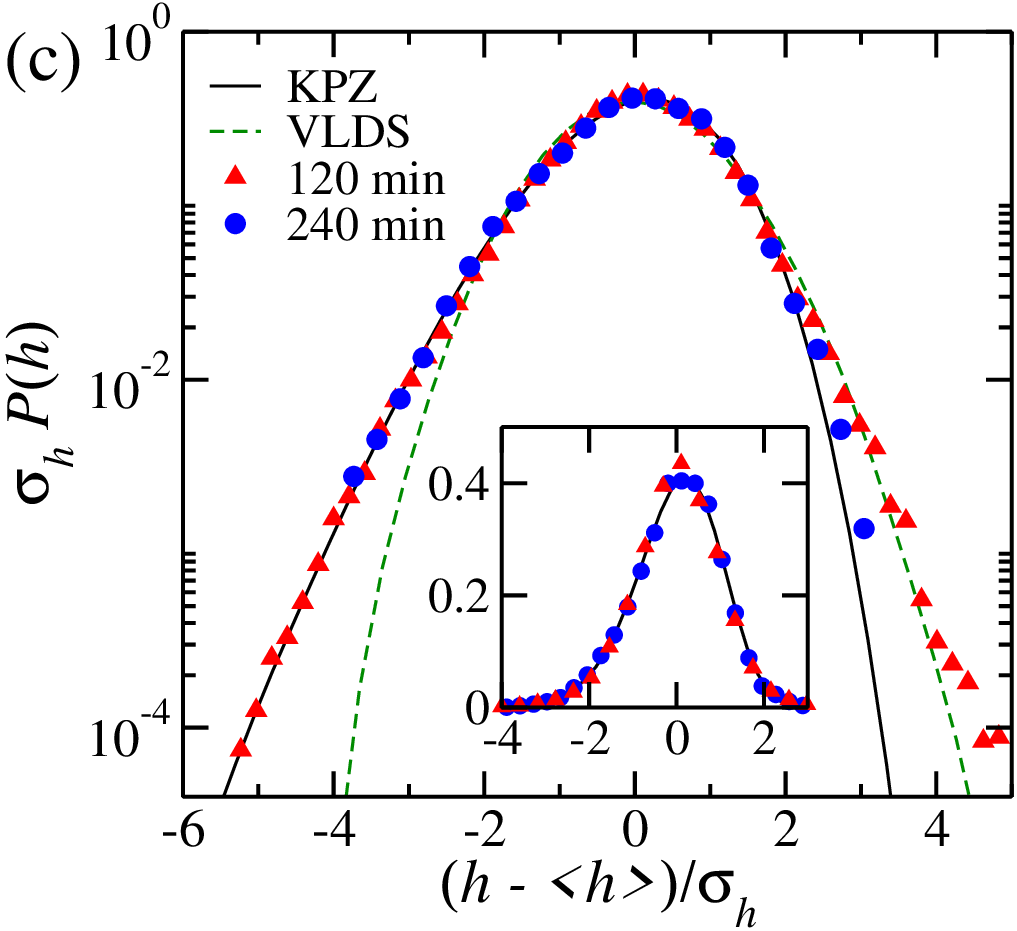}
\includegraphics*[width=4.20cm]{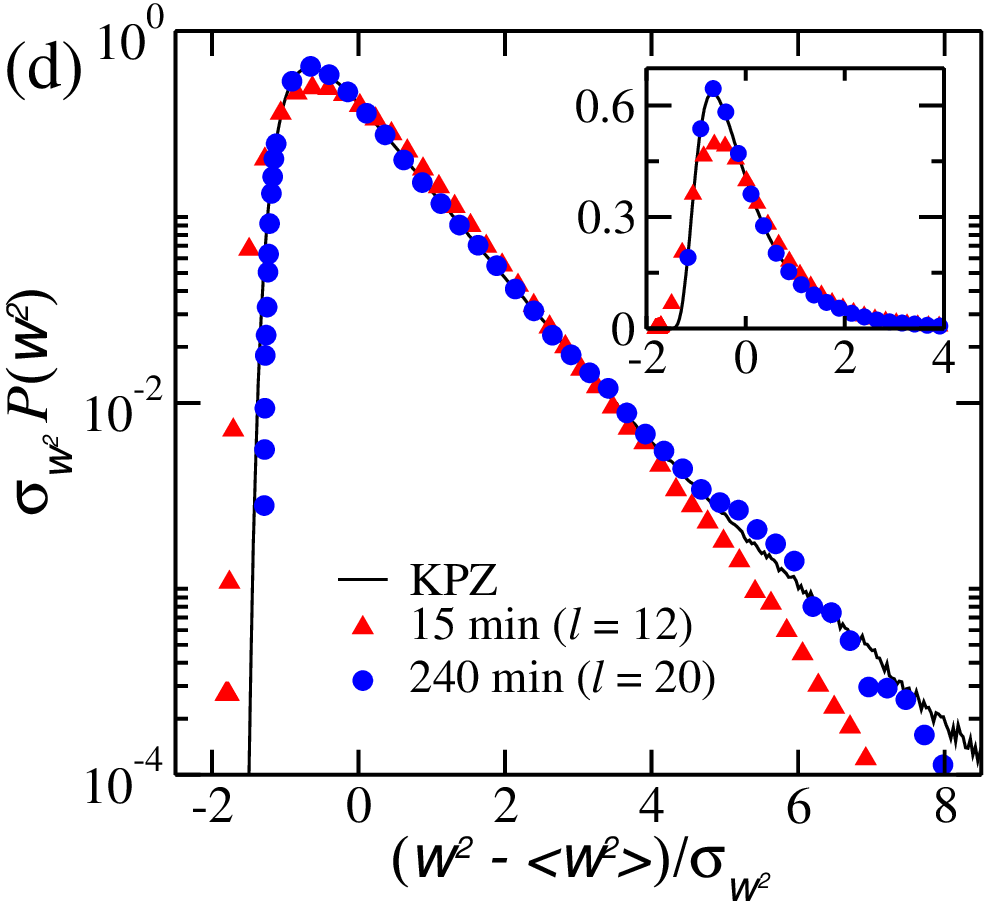}
\includegraphics*[width=4.20cm]{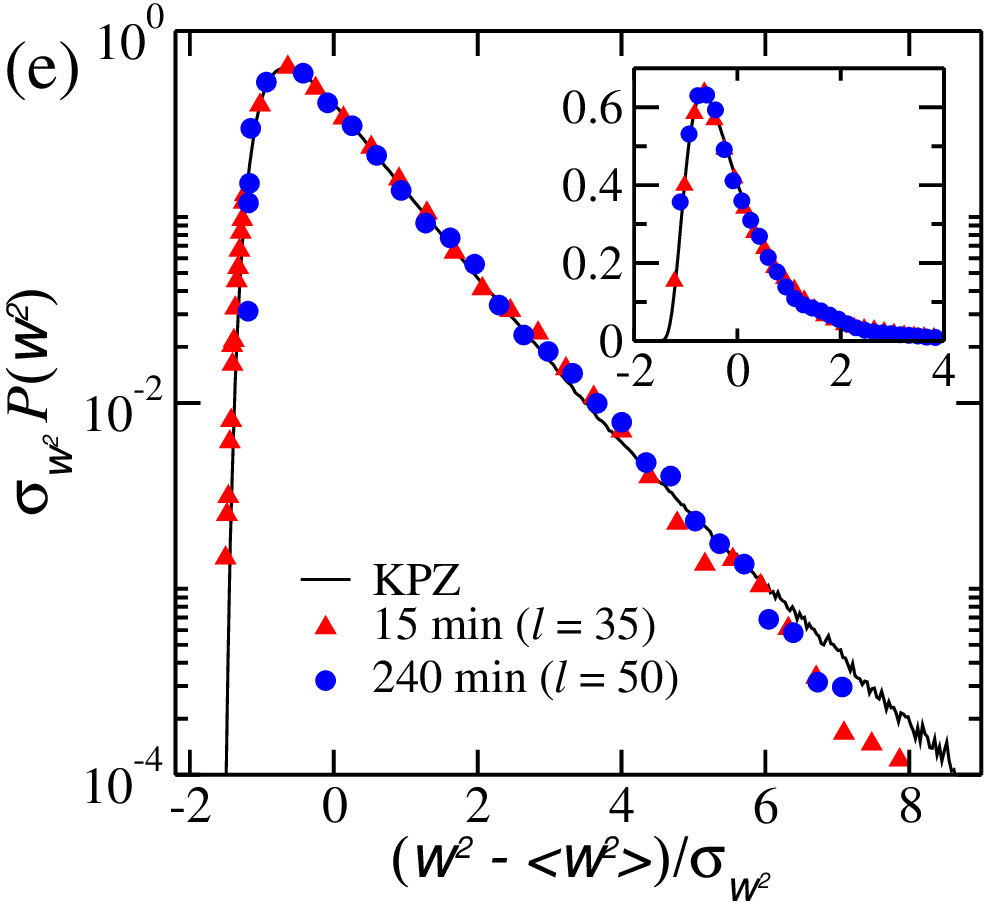}
\includegraphics*[width=4.20cm]{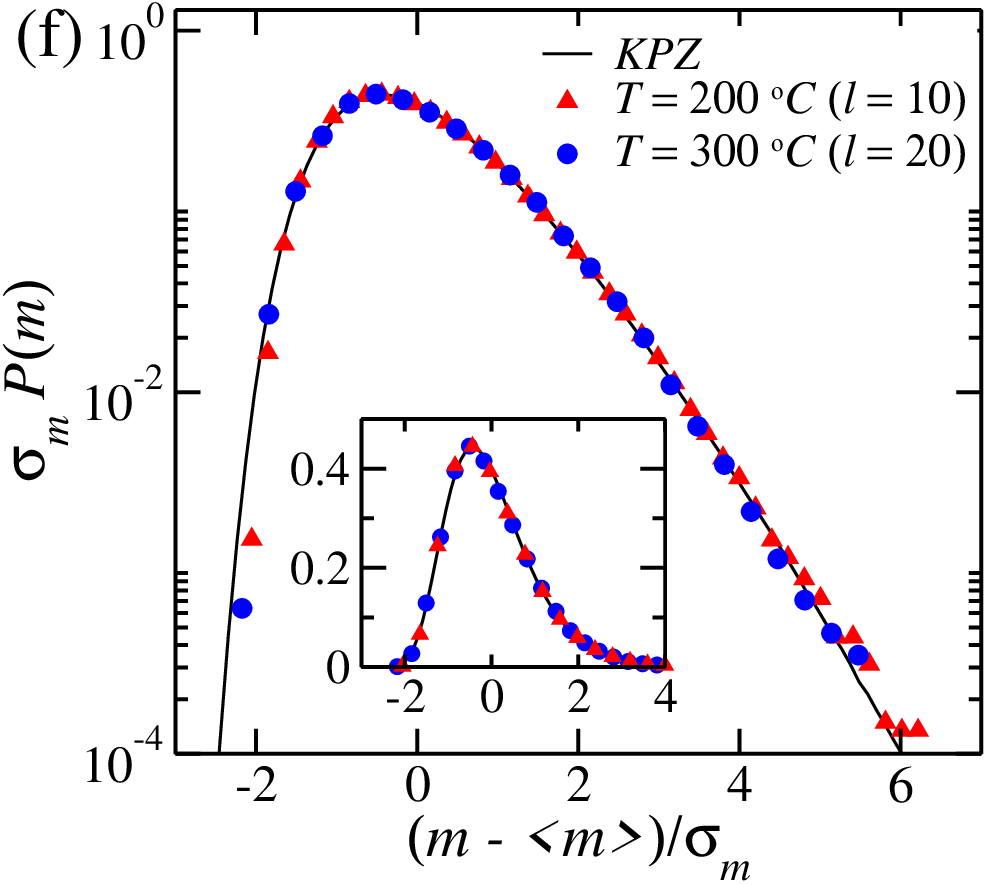}
\caption{Rescaled HDs for films grown at (a) $T = 150$, (b) $200$ and (c) $300\,^{\circ}\mathrm{C}$. Rescaled 
SLRDs for films grown at (d) $T=200$ and (e) $300\,^{\circ}\mathrm{C}$. (f) Rescaled MRHDs for large deposition 
times. Here, $\sigma_{X} \equiv \sqrt{\left\langle X^2\right\rangle - \left\langle X \right\rangle^2 }$. Insets 
display the same data of main plots in linear scale.}
\label{fig5}
\end{figure}

Figures \ref{fig5}a-c show the HDs for all $T$ studied. When $T=150\,^{\circ}\mathrm{C}$, experimental HDs are well-described by a Gaussian, as expected for an uncorrelated growth. For $T=200\,^{\circ}\mathrm{C}$, at short times, the HDs are also close to a Gaussian but, for large $t$, a nice collapse with the KPZ distribution is found. This agreement is confirmed by the skewness $S = 0.43(5)$ and kurtosis $K = 0.5(2)$ of the HDs, very 
close to the KPZ values $S=0.42(1)$ and $K=0.34(2)$ \cite{Halpin,tiago2d}. These results give further evidence of a
random-to-KPZ crossover, a subject of wide theoretical interest (see \cite{Juvenil} and references therein). Finally, for $T=300\,^{\circ}\mathrm{C}$ one still finds a reasonable agreement between the experimental HDs and the KPZ one, but now with a heavier left tail than the right one, yielding a negative skewness $S = -0.2(2)$, with kurtosis $K = 0.3(2)$. These values are consistent with KPZ class (with $\lambda < 0$ in Eq. \ref{eqKPZ}) within the error bars. Moreover, the HDs in Fig. \ref{fig5}c discard VLDS as the possible asymptotic UC.

{The experimental SLRDs for $T=200\,^{\circ}\mathrm{C}$ (Fig. \ref{fig5}d) also deviate from the KPZ distribution at short times, but have a nice agreement for long $t$, giving a final confirmation of a crossover towards KPZ.} For higher $T$, the SLRDs exhibit a nice collapse with the KPZ one (see Fig. \ref{fig5}e). We must remark that the stretched exponential decay in SLRDs right tail is a hallmark of the KPZ class and contrasts with the Gaussian decay of the VLDS distribution. Finally, the MRHDs for $T=200$ and $300\,^{\circ}\mathrm{C}$ are presented in Fig. \ref{fig5}f providing additional proof that {CdTe grows according to KPZ equation}.

The relation between local and coarse-grained dynamics can be understood as follows. For $T=150\,^{\circ}\mathrm{C}$, the low diffusion {and the energy barrier at GBs prevents coalescence/packing of grains and, thus, the propagation of correlations at interface}, so that inter-grain fluctuations evolve uncorrelated. This also happens at short times, for $T=200\,^{\circ}\mathrm{C}$, but the relaxation {process at GBs gives rise to an asymptotic correlated growth}. For higher $T$, where diffusion is more operative, these processes start early as well as the KPZ scaling. In terms of KPZ equation (Eq. \ref{eqKPZ}), the random growth at low $T$ implies $\nu \approx 0$ and $\lambda \approx 0$. For $T=200\,^{\circ}\mathrm{C}$, one expects $\lambda>0$, but small, so that growth is dominated by noise initially and by non-linear effects asymptotically. The absence of a crossover when $T=250\,^{\circ}\mathrm{C}$ \cite{Renan} indicates a larger $\lambda>0$. Thus, $\lambda(T)$ 
seems to be a positive increasing function in this range of $T$. {As discussed in Ref. \cite{Renan}, the possible origin of this KPZ growth is the complex coalescence/packing dynamics of the polycrystalline grains, where some grains cover their neighbors. Due to shape constraints, they do not necessarily fills all available space in its neighborhood, producing a positive velocity excess ($\lambda>0$) in the growth, similar to the lateral aggregation in ballistic deposition \cite{barabasi}. This process is more operative as higher is $T$, due to the larger inter-grain diffusion, so larger should be $\lambda$.} Otherwise, for $T=300\,^{\circ}\mathrm{C}$, the negative skewed HDs reveals $\lambda<0$, which is typical of KPZ systems where there exists deposition refuse as, for example, in the RSOS model \cite{RSOS}. Therefore, a possible explanation for $\lambda<0$ is that the sticking coefficient is smaller in regions with very large slopes at surface. Indeed, one sees in Fig. \ref{fig2}b that $\left  \langle (\nabla h)^2 \right\rangle$ for $T=300\, ^{\circ}\mathrm{C}$ is {larger than} for lower $T$. This can explain why this effect appears only at high $T$ \cite{Cuerno,Zhao}. Anyway, it is astonishing that so contrasting KPZ mechanisms can emerge in CdTe growth and indicate the possibility of {control and even} turn off the non-linearity (i.e., to make $\lambda = 0$) by only adjusting $T$.

\section{Final remarks} 

We finish stressing that the detailed morphological analysis performed here is \textit{imperative} to determine the UC of the system. Since the complex mound evolution gives rise to finite-size corrections, crossover effects/transient anomalous scaling, {it is not possible to drawn any} conclusion about the asymptotic growth dynamic based only on the traditional study of the scaling exponents. This {should explain} why reliable experimental evidences of KPZ and other classes are so rare. Notwithstanding, we {show} that HDs, SLRDs and MRHDs are less susceptible to the above effects and have allowed us to determine, conclusively, that CdTe surface fluctuations for films grown at $T \in [200, 300 \,^{\circ}\mathrm{C}]$ evolve according to the KPZ equation. Therefore, rather than a complementary analysis, the study of distributions is a crucial tool to unveil the growth dynamics. We believe that this findings will motivate future works in the same vein, as well as the application of these methods in previously studied systems.

%


\acknowledgments
This work was supported by FAPEMIG, CAPES and CNPq (Brazilian agencies).



\begin{thebibliography}{0}

 
\bibitem{evansrev} J. W. Evans, P. A. Thiel, and M. C. Bartelt, Surf. Sci. Rep. {\bf 61}, 1 (2006).



\bibitem{solarcell} M. Ledinsky \textit{et al.}, , App. Phys. Lett. \textbf{105}, 111106, (2014).


\bibitem{spin} A. Aqeel, I. J. Vera-Marun, B. J. van Wees and T. T. M. Palstra, App. Phys. Lett. \textbf{116}, 153705, (2014).


\bibitem{contact} G. Fisichella, G. Greco, F. Roccaforte and F. Giannazzo, App. Phys. Lett. \textbf{105}, 063117, (2014).


\bibitem{barabasi} A.-L. Barabasi and H. E. Stanley, \textit{Fractal Concepts in Surface Growth} (Cambridge University Press, Cambridge, UK, 1995).

\bibitem{Krugrev} J. Krug, Adv. in Phys. \textbf{46}, 139 (1997).

\bibitem{Marco} J. Maunuksela \textit{et al.}, Phys. Rev. Lett. \textbf{79}, 1515 (1997).

\bibitem{Takeuchi} K. A. Takeuchi, M. Sano, Phys. Rev. Lett. \textbf{104}, 230601 (2010); K. A. Takeuchi, M. Sano, T. Sasamoto and H. Spohn, Sci. Rep. \textbf{1}, 34 (2011).

\bibitem{Yunker} P. J. Yunker \textit{et al.}, Phys. Rev. Lett. {\bf 110}, 035501 (2013).

\bibitem{Cuerno} F. Ojeda, R. Cuerno, R. Salvarezza, and L. V\'azquez, Phys. Rev. Lett. \textbf{84}, 3125 (2000).

\bibitem{Renan} R. A. L. Almeida, S. O. Ferreira, T. J. Oliveira, and F. D. A. Aarao Reis, Phys. Rev. B \textbf{89}, 045309 (2014).

\bibitem{HHExper} T. Halpin-Healy, and G. Palasantzas, Europhys. Lett. 105, 50001 (2014).

\bibitem{KPZ} M. Kardar, G. Parisi, and Y.-C. Zhang, Phys. Rev. Lett. \textbf{56}, 889 (1986).


\bibitem{Ferreira} S. O. Ferreira \textit{et al.}, Appl. Phys. Lett. \textbf{88}, 244102 (2006); F. S. Nascimento \textit{et al.}, Europhy. Lett. \textbf{94}, 68002 (2011).
%
\bibitem{Hamouda} A. B. H. Hamouda, A. Pimpinelli, and R. J. Phaneuf, Surf. Sci. {\bf 602}, 2819 (2008).
%
\bibitem{Huo} S. Huo, W. Schwarzacher, Phys. Rev. Lett. \textbf{86}, 256 (2001); M. C. Lafouresse, P. J. Heard and 
 W. Schwarzacher, Phys. Rev. Lett. \textbf{98}, 236101 (2007).

\bibitem{CLi} C. Li \textit{et al.}, Phys. Rev. Lett. \textbf{112}, 156103 (2014).
 
\bibitem{Saba} M. Saba \textit{et al.}, Nature (London) \textbf{414}, 731 (2001).

\bibitem{Manna} L. Manna \textit{et al.}, Nature mat. \textbf{2}, 382 (2003).

\bibitem{Otero} A. L. -Otero, Thin Solid Films \textbf{49.1}, 3-57 (1978).
%
\bibitem{Suela} J. Suela \textit{et al.}, J. Appl. Phys. \textbf{107}, 064305 (2010).
%


\bibitem{Igor} I. R. B. Ribeiro \textit{et al.}, J. Phys. D: Appl. Phys. \textbf{40}, 4610 (2007).
%
\bibitem{tiago3} T. J. Oliveira and F. D. A. Aarao Reis, J. Appl. Phys. \textbf{101}, 063507 (2007); Phys. Rev. E \textbf{83}, 041608 (2011).
%
\bibitem{Anom} J. M. L\'opez, Phys. Rev. Lett. \textbf{83}, 4594 (1999); J. J. Ramasco, J. M. L\'opez, M. A. Rodr\'iguez, Phys. Rev. Lett. \textbf{84}, 2199 (2000).
%
\bibitem{FVS} F. Family and T. Vicsek, J. Phys. A \textbf{18}, L75 (1985). 
%
\bibitem{Sukarno} S. O. Ferreira \textit{et al.}, J. Appl. Phys. \textbf{93}, 1195 (2003).
%
\bibitem{Tello_Gonzales} J. S. Tello, A. F. Bower, E. Chason, and B. W. Sheldon, Phys. Rev. Lett. \textbf{98}, 216104 (2007); A. Gonz\'{a}lez-Gonz\'{a}lez, C. Polop, E. Vasco, Phys. Rev. Lett. \textbf{110}, 056101 (2013).
%
\bibitem{Kwon} D. Kwon \textit{et al.}, J. Appl. Phys. \textbf{116}, 183501 (2014).
%
\bibitem{Sivananthan} L. A. Almeida \textit{et al.}, J. Elect. Mat. \textbf{25}, 1402 (1996).
%
\bibitem{CRSOS} Y. Kim, D. K. Park, and J. M. Kim, J. Phys. A \textbf{27}, L533 (1994); F. D. A. A. Reis, Phys. Rev. E \textbf{70}, 031607 (2004).
%
\bibitem{Siniscalco} D. Siniscalco, M. Edely, J.-F. Bardeau, and N. Delorme, Langmuir \textbf{29}, 717 (2013).
%
\bibitem{VLDS} J. Villain, J. Phys. I \textbf{1}, 19 (1991); Z.-W. Lai and S. Das Sarma, Phys. Rev. Lett. \textbf{66}, 2348 (1991).
%
\bibitem{HDsteor} For reviews on theorectical developments in the KPZ class see, e. g., T. Kriecherbauer and J. Krug, J. Phys. A 43, 403001 (2010); I. Corwin, Random Matrices Theory Appl. 1, 1130001 (2012).
%
%
\bibitem{Halpin} T. Halpin-Healy, Phys. Rev. Lett. \textbf{109}, 170602 (2012); Phys. Rev. E \textbf{88}, 042118 (2013).
%
\bibitem{tiago2d} T. J. Oliveira, S. G. Alves and S. C. Ferreira, Phys. Rev. E \textbf{87}, 040102(R) (2013).
%
\bibitem{SLRDs} G. Foltin, K. Oerding, Z. R\'acz, R. L. Workman, and R. K. P. Zia, Phys. Rev. E \textbf{50}, R639 (1994); Z. R\'acz and M. Plischke, Phys. Rev. E \textbf{50}, 3530 (1994); F. D. A. Aarao Reis, Phys. Rev. E \textbf{72}, 032601 (2005).

\bibitem{Paiva} T. Paiva and F. D. A. A. Reis, Surf. Sci. \textbf{601}, 419 (2007).
%
\bibitem{MRHDs} S. Raychaudhuri, \textit{et. al}, Phys. Rev. Lett. \textbf{87}, 136101 (2001); S. N. Majumdar and A. Comtet, Phys. Rev. Lett. \textbf{92}, 225501 (2004); D.-S. Lee, Phys. Rev. Lett. \textbf{95}, 150601 (2005); T. J. Oliveira and F. D. A. Aarao Reis, Phys. Rev. E \textbf{77}, 041605 (2008).
%
%
 \bibitem{Juvenil} J. S. Oliveira Filho, T. J. Oliveira, and J. A. Redinz, Physica A \textbf{392}, 2479 (2013). 
%
%
\bibitem{RSOS} J. M. Kim and J. M. Kosterlitz, Phys. Rev. Lett. \textbf{62}, 2289 (1989).
%

\bibitem{Zhao} Y.-P. Zhao, J. T. Drotar, G.-C. Wang, and T.-M. Lu, Phys. Rev. Lett. \textbf{87}, 136102 (2001).



\end{thebibliography}
\end{document}